\documentclass[conference]{IEEEtran}
\usepackage{amsfonts}
\IEEEoverridecommandlockouts

\ifCLASSINFOpdf
\else
\fi

\usepackage{epsfig}
\usepackage{graphicx}
\usepackage{psfig}
\usepackage{subfigure}
\usepackage{epsf}
\usepackage[cmex10]{amsmath}
\usepackage{booktabs}
\usepackage{fancyhdr}

\newcommand{\bc}{\begin{center}}
	\newcommand{\ec}{\end{center}}
\newcommand{\be}{\begin{equation}}
\newcommand{\ee}{\end{equation}}
\newcommand{\bea}{\begin{eqnarray}}
\newcommand{\eea}{\end{eqnarray}}

\begin{document}
	\title{Intelligent Resource Allocations for IRS-Assisted OFDM Communications: A Hybrid MDQN-DDPG Approach}
	\author{Wei~Wu$^*$, Fengchun~Yang$^*$, Fuhui~Zhou\textsuperscript{\dag}, Han~Hu$^*$, Qihui~Wu,\textsuperscript{\dag}
and Rose Qingyang Hu\textsuperscript{\ddag}\\
		$^*$Nanjing University of Posts and Telecommunications, Nanjing, China,\\
		\textsuperscript{\dag}Nanjing University of Aeronautics and Astronautics, Nanjing, China,\\
		\textsuperscript{\ddag}Utah State University, USA.\\
		Emails: \{weiwu, 1020010608\}@njupt.edu.cn, {zhoufuhui}@ieee.org, \\
		\ han\_h@njupt.edu.cn, wuqihui@nuaa.edu.cn, rose.hu@usu.edu
\thanks{This work was supported by the National Key R\&D Program of China under Grant 2020YFB1807602, the National Natural Science Foundation of China under Grant 61901231, under Grant 62071223, under Grant 62031012, and Grant 61931011, the National Key Scientific Instrument and Equipment Development Project under Grant 61827801, the China Postdoctoral Science Foundation under Grant 2020M671480 and Postdoctoral Science Foundation of Jiangsu (2020Z295),  young Elite Scientist Sponsorship Program by CAST.}	
	}
	\maketitle
	
	\IEEEpeerreviewmaketitle
	\begin{abstract}
	In this paper, we study the resource allocation problem for an intelligent reflecting surface (IRS)-assisted OFDM system. The system sum rate maximization framework is formulated by jointly optimizing subcarrier allocation, base station transmit beamforming and IRS phase shift. Considering the continuous and discrete hybrid action space characteristics of the optimization variables, we propose an efficient resource allocation algorithm combining multiple deep Q networks (MDQN) and deep deterministic policy-gradient (DDPG) to deal with this issue. In our algorithm, MDQN are employed to solve the problem of large discrete action space, while DDPG is introduced to tackle the continuous action allocation. Compared with the traditional approaches, our proposed MDQN-DDPG based algorithm has the advantage of continuous behavior improvement through learning from the environment. Simulation results demonstrate superior performance of our design in terms of system sum rate compared with the benchmark schemes. 
\end{abstract}
\begin{IEEEkeywords}
	OFDM, intelligent reflecting surface, beamforming matrix, phase shift matrix, channel assignment, hybrid action space, MDQN-DDPG.
\end{IEEEkeywords}
\IEEEpeerreviewmaketitle
\section{Introduction}
Orthogonal frequency division multiplexing (OFDM) is  widely used in many communication systems such as LTE and fifth generation  wireless communication networks \cite{rq1}.
By utilizing orthogonal subcarriers, it can achieve high-speed and robust information transmission, and effectively avoid inter-channel interference \cite{w1}.
Moreover, the system performance can be significantly improved by optimizing the channel assignment and power control \cite{tj4}.
With the rapid development of mobile internet and wireless services, we are facing the explosive growth of mobile data and higher data rate requirements.
However, the wireless channel fading significantly decreases the performance of the OFDM communication system and the user experience.
Therefore, how to improve the performance of the OFDM communication system has become an urgent common concern in both industry and academia \cite{6g}.

Recently, intellligent reflecting surface (IRS) has been proposed as a promising solution to enhancing the signal quality at the desired receiver in a cost-effective and energy-efficient way \cite{ee1}.
Specifically, IRS is a reflective array composed of a large number of low energy consuming and low-cost passive reflective elements \cite{ee3}.
Each element can independently adjust the phase shift of the incident signal \cite{hj1}, so as to cooperatively change the propagation of the reflected signal in order to achieve the desired channel response.
By properly adjusting the phase shift of IRS elements, the reflected signals of different paths can be coherently combined at the receiver to maximize the achievable rate of the link.
Therefore, with such a characteristic, IRS is able to overcome the problem of OFDM channel fading, and provide higher data rate for the network and better experience for the users.
The authors in \cite{zc1} studied the scenario of IRS assisted OFDM communications.
However, the authors only considered the single antenna BS scenario.
The authors applied alternating optimization and successive convex approximation (SCA) techniques to solve the optimization problems of joint IRS reflection coefficient, time-frequency resource block allocation and power allocation, so as to maximize the common (minimum) rate among all users.
In \cite{yw1}, the authors studied the adaptive transmission scenario of an IRS assisted uplink OFDM system, and an algorithm based on semidefinite relaxation technology was employed to improve the average achieveable rate.

However, most of the above work adopted traditional mathematical skills such as alternating optimization and successive convex approximation. 
It is difficult for these complex mathematical operations and numerical optimization methods are difficult to meet the real time processing requirements of large-scale heterogeneous communication systems.
Recently, deep reinforcement learning (DRL) has attracted wide attention from researchers due to its real time performance \cite{tj1}-\cite{tj3}.
In \cite{zc2}, DRL was used to solve the joint optimization problem of beamforming matrix and IRS phase shift matrix.
In \cite{ee1}, the authors proposed a secure beamforming method based on DRL.
The authors in \cite{irs1} used DRL to optimize the IRS phase shift matrix. 
Simulation results showed that the DRL algorithm can achieve the upper bound of system performance using lower time consumption compared to the positive semidefinite relaxation algorithm.
To the authors' best knowledge, in the IRS-assisted OFDM resource allocation scenario, there has been no relevant research on using DRL to optimize channel allocation, beamforming matrix and IRS passive beam phase shift, which is of crucial importance for making full use of IRS to improve the performance of OFDM systems. 

In this paper, we study an IRS assisted OFDM communication system. 
Our goal is to achieve the maximum total system rate while ensuring the minimum transmission rate requirements of users.
The optimization problem is transformed into a Markov decision process. 
Aiming at tackling the problem of discrete and continuous hybrid action space, we propose an efficient resource allocation algorithm based on multiple deep Q networks and deep deterministic policy-gradient to jointly optimize channel allocation, beamforming and IRS passive beam phase shift.
Simulation results show that the proposed algorithm can significantly improve the sum rate of the system, and quickly converge.

The rest of the paper is organized as follows.
Section II presents the system model. Section III presents the resource allocation algorithm based on MDQN-DDPG. Section IV gives simulation results. Section V presents the conclusion.
\section{System Model}
We consider an IRS-assisted downlink OFDM communications system as shown in Fig. 1, in which IRS is used to enhance the signal strength between base station (BS) and users.
The BS is equipped with $M$ antennas and the user is equipped with a single antenna. The IRS contains $N$ passive reflection elements, and is connected to a controller that adjusts the IRS mode for the required signal reflection.
${\cal K} = \left\{ {1, \cdots ,K} \right\}$ and  ${\cal C} = \left\{ {1, \cdots ,C} \right\}$ represent the user set and channel set, respectively.
	\begin{figure}[h]
	\centering
	\includegraphics[width=2.62in]{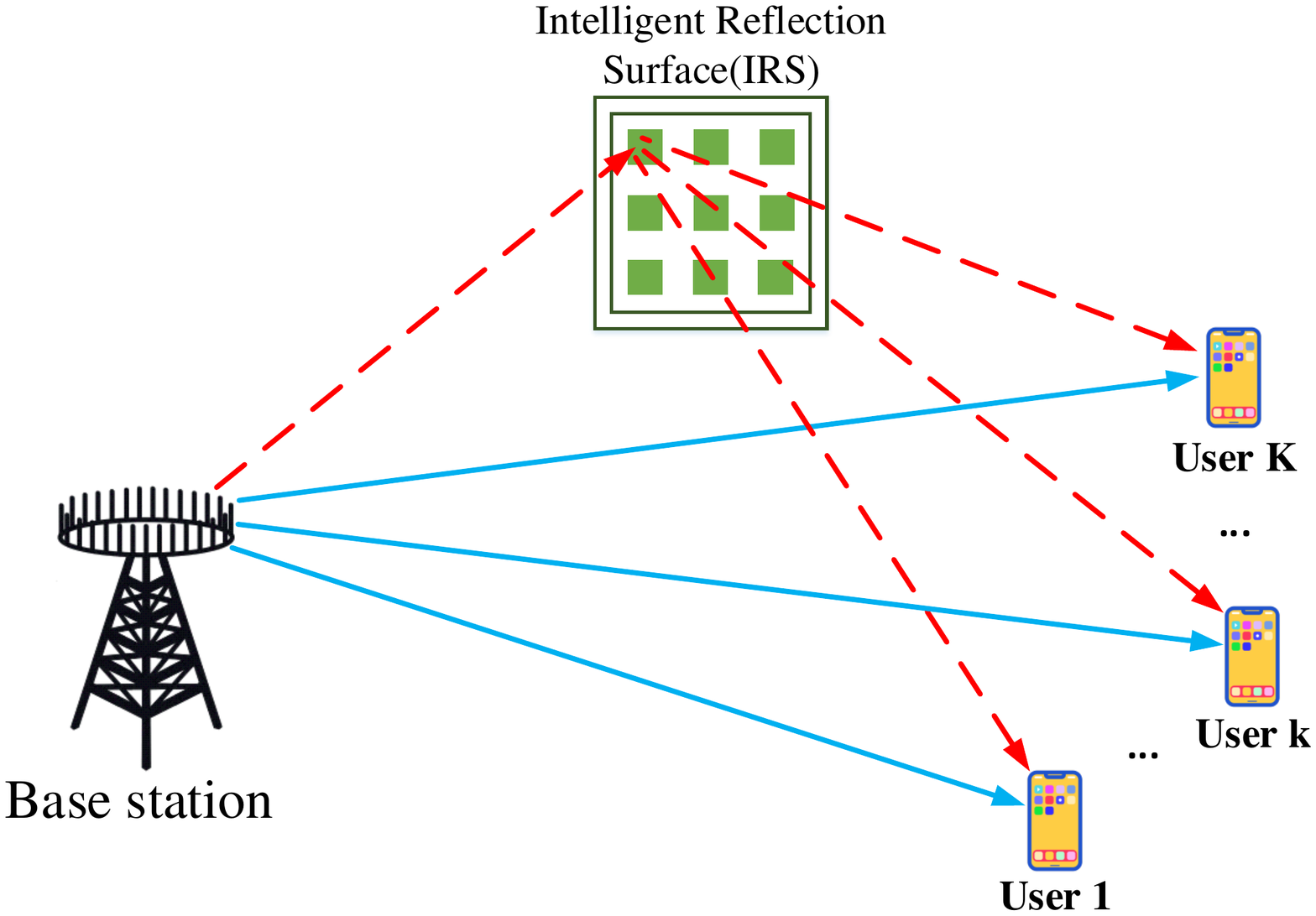}	\label{model}
	\caption{An IRS-assisted downlink OFDM communications system with multiple users.}
\end{figure}

Each user can receive signals from the BS through both direct and reflective links.
$\mathbf H_{B,R} \in \mathbb{C}^{M \times N \times C }$, $\mathbf h_{R,k} \in \mathbb{C}^{N \times 1 \times C }$ and $\mathbf h_{B,k} \in \mathbb{C}^{M \times 1 \times C }$ denote the channel coefficients from the BS to the IRS, from the IRS to the user $k$ and from the BS to user $k$, respectively. 
Let $\mathbf{\Theta} = \text{diag}({\Theta_{1}}, {\Theta_{2}}, \cdots, {\Theta_{N}}) \in \mathbb{C}^{N \times N }$ denote the phase shift matrix related to IRS, where {$\Theta_{n}=w_ne^{j\theta_n}$} comprises amplitude correlation coefficient $w_n \in \left[{0,1}\right]$ and phase correlation coefficient $\theta_{n} \in \left[ {0,2 \pi} \right]$.
Since each element of the IRS is designed to do complete reflection, the amplitude correlation coefficient is set as $w_n=1$ [9].
At the BS, the beamforming vector for user $k$ on the  channel $c$ is denoted as $\mathbf{f}_k^c$.
The total transmit power at the BS is constrained as $\sum_{k=1}^K{\sum_{c=1}^C\rho_k^c||\mathbf{f}_k^c||^2} \leq P_T$, where $P_T$ is the maximum transmission power of BS.
$\rho _k^c$ indicates the user's channel usage. If user $k$ uses channel $c$, then $\rho_k^c=1$; otherwise, $\rho_k^c=0$.

Based on the above descriptions, the signal received at user $k$ can be given as
\be 
{y_k} = \sum\limits_{c = 1}^C {\rho _k^c\left( {{{\bf{h}}_{B,k}^c}^T + {{\bf{h}}_{R,k}^c}^T{\bf{\Theta }}{{\bf{H}}_{B,R} ^c}} \right)} {\bf{f}}_k^c{s_k} + \sigma _k^2,
\ee
where $\sigma_k^2$ is additive complex Gaussian random noise at the user $k$,
$s_k$ is the emission symbol of the user $k$ satisfying $\mathbb{E}\left[ {{{\left| {{s_k}} \right|}^2}} \right] = 1$.
${\bf{h}}_{B,k}^c$ represents the channel coefficient between the BS and user $k$ when the BS transmits data to user $k$ using channel $c$.
Similarly, when the channel $c$ is used for transmission, ${\bf{H}}_{B,R} ^c$ and ${\bf{h}}_{R,k}^c$ represent the channel coefficients between BS to IRS and IRS to user $k$, respectively.

Based on (1), the transmission rate from BS to the user $k$ on the $c$-th channel can be expressed as
\be
\ R_k^c = \frac{B}{C}\rho _k^c{\log _2}\left[ {1 + \frac{{{{\left| {\left( {{{\bf{h}}_{d,k}^c}^T + {{\bf{h}}_{R,k}^c}^T{\bf{\Theta }}{{\bf{H}}_{B,R}^c}} \right){\bf{f}}_k^c} \right|}^2}}}{{\sigma _k^2}}} \right],
\ee
where $B$ denotes the bandwidth.

Therefore, the total transmission rate of all the $K$ users can be given as
\be
R =\sum_{k=1}^K{\sum_{c=1}^CR_k^c} .
\ee

Our design goal is to find the optimal $\mathbf{f}_k^c$,$\mathbf {\Theta}$ and $\rho _k^c$ to maximize the system sum rate on the premise of meeting the minimum transmission rate requirements of users and BS transmit power constraint.
The optimization problem can be formulated as
\begin{subequations}
\begin{alignat}{5}
\ &\max_{\mathbf{\rho},\mathbf{f},\mathbf{\Theta}}~R \nonumber\\
s.t.~~&\sum_{k=1}^K{\sum_{c=1}^C\rho_k^c||\mathbf{f}_k^c||^2} \leq P_T,\\
&\ |\theta_n|=1,\forall n\in \mathcal{N},\\
&\rho_k^c\in \{0,1\},\forall k \in \mathcal{K},c\in \mathcal{C},\label{10c}\\
&\sum_{k=1}^K\rho_k^c\leq 1,\forall c\in \mathcal{C},\\
&\sum_{c=1}^CR_k^c\geq R_k^{min},\forall k\in \mathcal{K},
\end{alignat}
\end{subequations}
where $R_k^{min}$ represents the minimum transmission rate requirement of the user $k$.
The constraint (4a) ensures that the total transmit power of the BS is less than its maximum transmit power,
the constraint (4b) indicates that the reflection unit of IRS is the total reflection with the amplitude correlation coefficient being 1,
the constraints (4c) and (4d) enforce that a channel can only be occupied by one user instead of  multiple users,
the constraint (4e) means that each user must meet their own minimum transmission rate requirements.
Due to the joint optimization of discrete subcarrier allocation, continuous passive beam phase shift and beamforming, problem (4) is highly non-convex and difficult to solve. Therefore, a joint DQN and DDPG algorithm based on DRL is proposed to provide a global optimal solution.

\section{DRL Based  Resource Allocation Framework}
\subsection{Problem Formulation Based on MDP}
Model free RL is a dynamic decision-making tool, which can solve the decision-making problem by learning the optimal solution in dynamic environment [9]. 
We model the formulated discrete continuous optimization problem as an MDP problem.
The IRS-assisted wireless communication scenario is regarded as an environment, and the central controller of BS is regarded as an agent.
In addition to environment and agent, the MDP problem also includes state, action, reward and transition probability. 
The key elements of MDP are described as follows.

\textbf{State space}: Set $S$ as the state space. The system state $s^t\in S$ at current time includes channel allocation at the previous time, IRS passive beam phase shift, beamforming matrix, achievable rate of all users and channel vectors corresponding to $C$ channels. $s^t$ at t is defined as 
\be
s^t=\{a_1^{t-1},a_2^{t-1},\{R_k^{t-1}\}_{k\in \mathcal{K}},\{\mathbf{h}_{c}^t\}_{c\in \mathcal{C}}\}.
\ee
where $a_1^{t-1}$ and $a_2^{t-1}$ represent discrete action and continuous action at $t-1$ time respectively.
Since the input of neural network can only be real numbers instead of complex numbers, the channel state $\mathbf{h}_{c}^t$ is divided into real part and imaginary part as input into neural network, respectively.

\textbf{Action space}: Set $A$ as the action space, and the central controller of BS selects the optimal action according to the current environmental state. 
Since the optimization problem includes discrete and continuous hybrid actions, the action space can be divided into two parts. One part is for discrete actions including user's channel allocation $\lambda$, and the other part is for continuous actions including beamforming vector $\mathbf{f}$ of BS and passive beam phase shift $\mathbf{\Theta}$ of IRS. Hence, action $a\in A$ can be defined as
\be
a^t=\{a_1^t,a_2^t\},
\ee
where $a_1^t=\{\lambda_c\}_{c\in  \mathcal{C}}$ and $a_2^t=\{\{\mathbf{f}_c\}_{c\in \mathcal{C}},\mathbf{\Theta}\}$.

\textbf{Transition probability}: $Pr(s^{t+1}|s^t,a^t)$ is defined as the transition model, which is the probability of switching from state $s^t\in S$ to the new state $s^{t+1}\in S$ after taking action $a^t\in A$.

\textbf{Reward function}: The reward function is a very important part of DRL design process, which is closely related to the desired goal of the system.
According to the optimization problem formulated in Section II, the goal of the reward function contains two aspects. One is to maximize the system sum rate and the other one is to meet the minimum data rate requirements.
Based on the above objective design, the reward function can be expressed as
\be
r=w_1R_{all}+w_2\sum_{k=1}^K\delta_k,
\ee
where $w_1$ and $w_2$ are constant coefficients, $R_{all}$ is the system sum rate, $\delta_k$ is a penalty item to punish the user whose current transmission rate does not meet the minimum transmission rate requirements,
which can be given as
\be 
\delta_k=\begin{cases}
0,&R_k\geq R_k^{min}\\
-R_k,&0<R_k< R_k^{min}\\
-b,&R_k=0,
\end{cases}
\ee
where $R_k$ is the transmission rate of the user $k$, $b>0$ is a constant coefficient.

The goal of the agent is to find an optimal strategy to maximize the long-term reward. The cumulative discount reward can be denoted as
\be
R^t=\sum_{\tau =0}^\infty \gamma ^\tau r^{t+\tau +1},
\ee
where $\gamma \in [0,1)$ is the discount rate.

$Q_\pi (s,a)$ is used as the state action value function. Given the state $s^t$, action $a^t$ and reward $R^t$, the Q function can be expressed as
\be
{Q_\pi }\left( {{s^t},{a^t}} \right) = {E_\pi }\left[ {{R^t}\left| {{s^t} = s,} \right.{a^t} = a} \right].
\ee

Then, the Q function that satisfies the Bellmann equation is obtained as
\be
\begin{split}
&{Q_\pi }\left( {{s^t},{a^t}} \right) = {E_\pi }\left[ {{r^{t + 1}}\left| {{s^t} = s,{a^t} = a} \right.} \right] + \gamma \sum\limits_{{s^{t + 1}} \in S} {} \\
&P\left( {{s^{t + 1}}\left| {{s^t},{a^t}} \right.} \right)\left( {\sum\limits_{{a^{t + 1}} \in A} {\pi \left( {{s^{t + 1}},{a^{t + 1}}} \right){Q_\pi }\left( {{s^{t + 1}},{a^{t + 1}}} \right)} } \right).
\end{split}
\ee

The Q-learning algorithm is used to search for the optimal policies $\pi ^*$. From (11), the optimal Q function associated with the optimal policy can be expressed as
\be
\begin{split}
Q^* \left(s^t,a^t\right)=&r^{t+1}+\gamma \sum_{s^{t+1}\in S}P\left(s^{t+1}|s^t,a^t\right)\\
&\max_{a^{t+1}\in A} Q^* \left(s^{t+1},a^{t+1}\right).
\end{split}
\ee

The Bellmann equation can be solved in a recursive way, and iterating (12) yields the optimal Q function. Therefore,, the update procedure for the iterative solution can be denoted as
\be
\begin{split}
Q\left(s^t,a^t\right)\leftarrow &(1-\alpha)Q\left(s^t,a^t\right)\\
&+\alpha \left(r^{t+1}+\gamma \max_{a^{t+1}}Q^*\left(s^{t+1},a^{t+1}\right)\right),
\end{split}
\ee
where $\alpha \in (0,1]$ is the learning rate of the Q function updates.

\subsection{Resource Allocation With  MDQN-DDPG}

In this section, we introduce the MDQN-DDPG framework, as shown in Fig. 2, in which MDQN is used to process the discrete actions, while DDPG is used for continuous actions.

	\begin{figure}[h]
	\centering
	\includegraphics[width=3.2in]{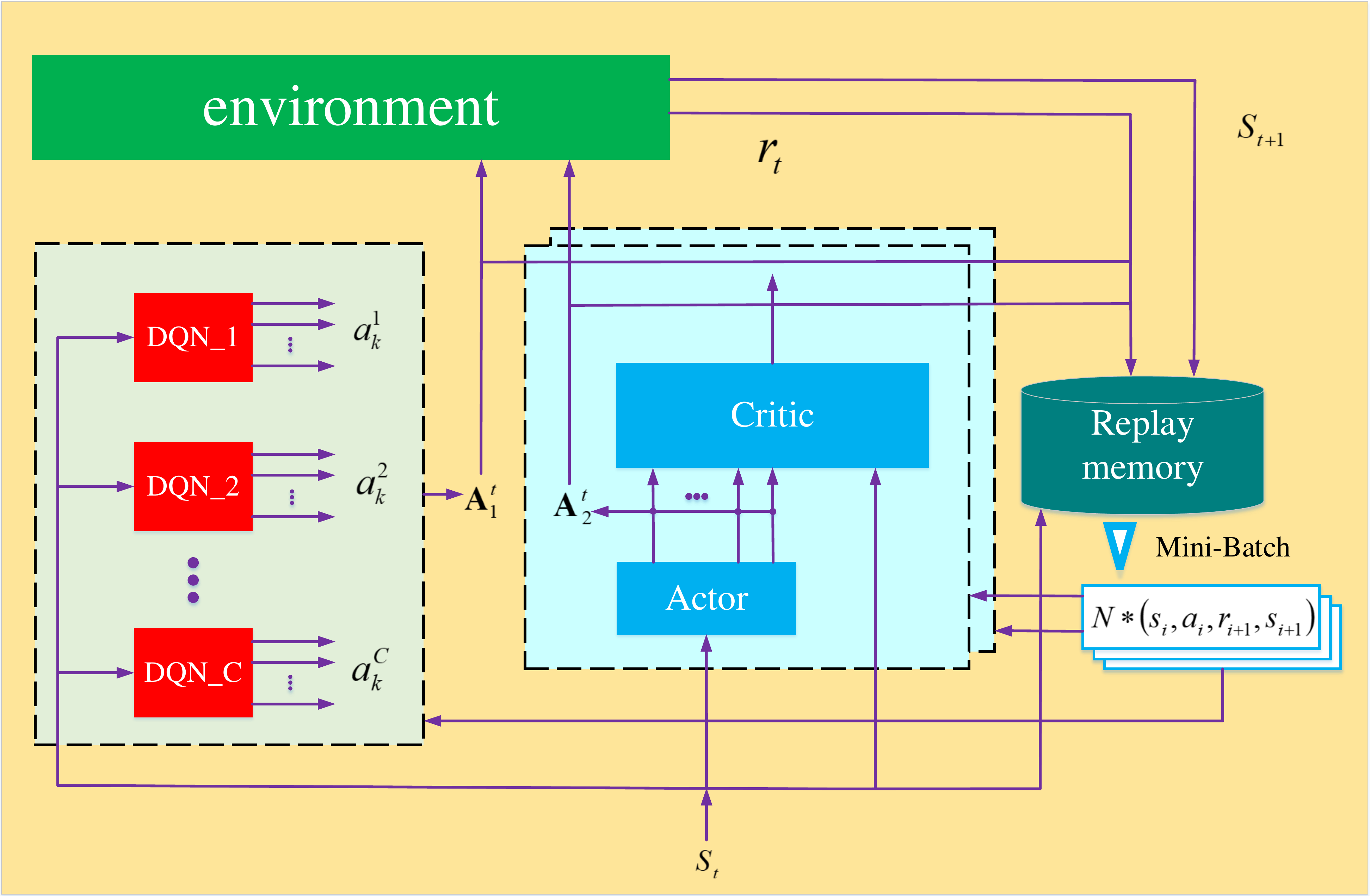}	\label{model}
	\caption{Proposed framework for resource allocation based on MDQN-DDPG.}
\end{figure}
\subsubsection{MDQN}
DQN is a classical method of DRL in many scenarios and can effectively deal with problems with large state space and large action space. 
One of its main features is that the output action is discrete.
Therefore, we adopt DQN framework to solve the channel allocation issue in our formulated problem, but if there is only one DQN, it will lead to huge action space. For example, in a system with $C$ channels, $K$ users, the action space reaches $K^C$.
Thus, we use multiple DQN for distributed processing, $C$ channels adopt $C$ DQN, DQN select the best user for the channel.  
In this way, the action space can be reduced from $K^C$ to $K*C$, to greatly improve the learning efficiency.

DQN is a Q network constructed by DNN to estimate the Q value of the selected action.
Specifically, with state S as network input, the Q network outputs the estimated Q values for all actions.
Following that, the $\epsilon$-greedy method can be adopted to carry out channel allocation to balance the exploration of action and the exploitation of known actions. In other words, it can randomly select one action among all the actions with $\epsilon$ probability or select action $a$ with the largest estimated value with 1-$\epsilon$ probability, which can be given as

\be
a=arg\max_{a\in A}Q\left(s,a;w\right),
\ee
where $0<\epsilon <1$. The DQN network can continue to select actions with high feedback rewards or explore actions that may have higher rewards but are not yet selected, so as to explore the entire action space and update Q values. $\omega$ represents the weight of the training network and leverages the experience replay buffer for continuously updating.

The random extraction of $N$ tuples $(s^j,a^j,r^{j+1},s^{j+1})$ from the experience replay buffer is to ensure the independence of the training tuples and avoid the problem caused by excessive correlation among the tuples.

The randomly sampled tuples $(s^j,a^j,r^{j+1},s^{j+1})$ can be used to generate target Q values
\be
y=r^{j+1}+\gamma \max_{a^{j+1}\in A}Q\left(s^{j+1},a^{j+1};\omega^-\right),
\ee
where $\omega^-$ represents the weight of the target network.
The loss function gives the difference between the predicted value of the neural network and the actual target value. It can be expressed as
\be
l(\omega)=\left(y-Q\left(s^j,a^j;w\right)\right)^2.
\ee
\subsubsection{DDPG}
For the problem with continuous actions, such as beamforming and IRS passive phase shifts, DQN does not work.
Different from the traditional method generating actions according to the probability, DDPG generates and determines actions according to parametric strategy. Moreover, it has neural network and Actor-Critic structure, which enables DDPG to deal with continuous action issue effectively.

The DDPG network contains four sub-networks, e.g,, the current Actor network composed of DNN for action selection, the current Critic network used to generate estimated Q value of the current action, as well as the target Actor network and target Critic network used to generate target value for training.
In summary, DDPG is an extension of actor-critic framework. It utilizes experience replay and double network methods to improve the convergence performance of the original framework.

For the Actor network, in order to balance the exploration of new actions and the exploitation of known actions, random noise is added to the output actions. 
After executing the action $a^i$, we can get the immediate reward $r^{i+1}$ and store the tuple $(s^i,a^i,r^{i+1},s^{i+1})$ in the experience replay buffer.
According to the DPG theorem, $N$ tuples can be randomly selected from the experience replay buffer to update the weight $\theta$ in the current actor network, and the update gradient is given as
\be
\nabla _\theta J(\mu)=\frac{1}{N}\sum_i\left[\nabla_aQ\left(s^i,a;\lambda\right)|_{a=\mu (s^i;\theta)}\nabla_\theta \mu \left(s^i;\theta\right)\right].
\ee

For the Critic network, we use the target Critic network and the target Actor network to update the weight $\lambda$.
The sampled tuples are used to generate target Q value of the current training, given as
\be
y=r^{i+1}+\gamma Q\left(s^{i+1},a;\lambda ^-\right)|_{a=\mu (s^{i+1};\theta ^-)},
\ee
where $\theta ^-$ and $\lambda ^-$ are the weights of the target Actor network and the target Critic network, respectively.

The loss function uses the difference between the predicted value and the target value, and the weight $\lambda$ can be updated by minimizing the loss function, that is
\be
l(\lambda)=\frac{1}{N}\sum_i\left[y-Q\left(s^i,a^i;\lambda\right)\right]^2.
\ee

In short, the whole process starts with extracting extract tuples from the experience replay buffer, followed by inputting the extracted tuples to the target network and the current network. Finally, we can solve the gradient using (17) and (19), and update the parameters $\theta$ and $\lambda$ corresponding to the current network. After a certain number of steps, DDPG copies the parameters of need to reword to the target network.
Algorithm 1 summarizes the details of MDQN-DDPG framework for IRS-assisted downlink OFDM system.
\begin{table}[htbp]
	\begin{center}
		\begin{tabular}{lcl}
			\\\toprule
			$\textbf{Algorithm 1}$: Resource Allocation in OFDM With MDQN-DDPG \\  \midrule
			\ \ \textbf{Input:} $\mathbf{H}_{BR},\mathbf{h}_{Rk},\mathbf h_{dk}$. Minimum transmission rate requirements\\
			\ \ for all users\\
			\ \ \textbf{Output:} optimal action $a=\{\mathbf{\rho},\mathbf{f},\mathbf{\Theta}\}$,$Q$ value function\\
			\ \ \textbf{Initialization:} experience replay buffer D with size D,\\
			\ \ the Q function of M DQNs, parameters $w$, parameters $w^-$,the \\
			\ \ parameters $\theta$ of the training Actor network, the parameter $\theta^-$ of the  \\
			\ \ target Actor network, the parameters $\lambda$ of the training Critic network \\
			\ \ and the parameters $\lambda^-$ of the target Critic network in the DDPG \\
			\ \ network, Channel allocation $\rho$, beamforming $\mathbf{f}$ and IRS phase shift $\mathbf{\Theta}$.\\
			\  1: \textbf{for} each episode \textbf{do}\\
			\  2:\ \ \ \  Collect $\mathbf{H}_{BR},\mathbf{h}_{Rk},\mathbf h_{dk}$ to observe an initial system state $s^0$;\\
			\  3:\ \ \ \ $\textbf{for}$ each step $t$ \textbf{do}\\
			\  4:\ \ \ \ \ \ $\textbf{for}$ each DQN agent $m$ \textbf{do}\\
			\  5:\ \ \ \ \ \ \ \ \ In the current state $s^t$, action $a_{1,m}^t$ is selected according to \\
			\  \ \ \ \ \ \ \ \ \ \ \ \ $\epsilon-$greedy policy;\\
			\  6:\ \ \ \ \ \ $\textbf{end for}$\\
			\  7:\ \ \ \ \ \ Obtain action $a_2^t$ from actor network;\\
			\  8:\ \ \ \ \ \ Execute action $a^t=\{a_1^t,a_2^t\}$ to obtain instant reward $r^t$ and\\
			\ \ \ \ \ \ \ \ \  next state $s^{t+1}$;\\
			\ 9:\ \ \ \ \ \ Store $(s^t,a^t,r^{t+1},s^{t+1})$ in the replay memory D;\\
			\ 10:\ \ \  \ \ Mini-batch of samples with size $N$ is randomly selected from the \\
			\ \ \ \ \ \ \ \ \ \ replay memory D;\\
			\ 11:\ \ \ \ \ $\textbf{for}$ each DQN agent $m$ \textbf{do}\\
			\  12:\ \ \ \ \ \ \ \ The SGD method is used to minimize the error between \\
			\ \ \ \ \ \ \ \ \ \ \ \ \ the predicted value and the target value, as shown in (16)\\
			\  13:\ \ \ \ \ \ \ \ Update DQN network parameters $w$.\\
			\  14:\ \ \ \ \ $\textbf{end for}$\\
			\  15:\ \ \ \ \ The loss function of the training Critic network represented by\\
			\ \ \ \ \ \ \ \ \ (19) is generated.\\
			\  16:\ \ \ \ \ Generating gradient $\nabla_aQ(s^t,a;\lambda)$ of training Critic network;\\
			\  17:\ \ \ \ \ Generating gradient $\nabla_\theta \mu (s^t;\theta)$ of training Actor network;\\
			\  18:\ \ \ \ \ Update training Actor network parameters $\theta$;\\
			\  19:\ \ \ \ \ Update training Critic network parameters $\lambda$;\\
			\  20:\ \ \ \ \ Update DQN target network parameters $w^-$ every $P$ steps;\\
			\  20:\ \ \ \ \ Update target Actor network parameters $\theta^-$ every $P$ steps;\\
			\  21:\ \ \ \ \ Update target Critic network parameters $\lambda^-$ every $P$ steps;\\
			\  22:\ \ \ $\textbf{end for}$\\
			\  23: \textbf{end for}.\\
			\bottomrule
		\end{tabular}
	\end{center}
\end{table}
\section{Simulation Results}
In this section, the performance of the proposed resource allocation algorithm is evaluated and compared with the benchmark schemes. In the simulation, the channel from BS to user is assumed to be Rayleigh fading, while the BS to IRS channel and the IRS to user channel are modeled as Rician fading.
According to \cite{ee1}, we can express the corresponding path fading as $PL=(PL_0-10\tau \log_{10}(d/D_0))$ dB, where $PL_0$ = 30 dB path loss with reference distance $D_0$ = 1 $m$.
We set the path loss exponent from BS to user as $\tau _{bu}=3.75$, and the path loss exponents from BS to IRS and IRS to user are $\tau _{br}=2.2$ and $\tau _{ru}=2.2$, respectively.

A three-dimensional coordinate system is established, $K$ single antenna ground users are randomly located in a 100 $m$ $*$ 100 $m$ rectangular area, and the lower left corner of the rectangular area is marked as $(100,0,0)$, while the upper right corner is marked as $(200,100,0)$. 
The BS and IRS are located at $(0,0,30)$ and $(75,100,50)$, respectively.
The background noise power of all users is $-169$ dBm. The number of antennas corresponding to BS is $M$ = 6, the number of users is $K$ = 3, the number of IRS phase shift units $N$ ranges from 16 to 64, and the transmission power range of BS is $15$ dBm to $40$ dBm.
In the proposed DRL based framework, DQN contains three hidden layers, while DDPG Actor network and Critic network contains two hidden layers. The learning rate of DQN is set as $0.002$, the learning rates of Actor network and Critic network are set to $0.001$ and $0.002$, respectively. The discount factor is set as $\gamma =0.99$, and the experience replay buffer is $D=6000$.

Fig. 3 shows the reward versus the iteration of the algorithm when $P_{T}=35$ dBm and $N$ = 16. For comparison, two existing schemes without IRS and with random selection of variable values are introduced. It is found that using the proposed MDQN-DDPG algorithm, in both without IRS-assisted and IRS-assisted cases, the rewards can be continuously improved and converge to a constant value at about 75 episodes.
Their early stage rewards are much lower because the user's transmission requirements are not satisfied.
Moreover, higher rewards can be achieved by our proposed IRS-assisted scheme than the scheme without IRS. This means that the employing IRS is conducive to significantly improve the system sum rate.
From Fig. 3, it is seen that the proposed MDQN-DDPG algorithm has greater reward than the random selection method. It is proved that the algorithm is effective and can obtain a better solution to the joint optimization problem.
\begin{figure}[h]
	\centering
	\includegraphics[width=3.0in]{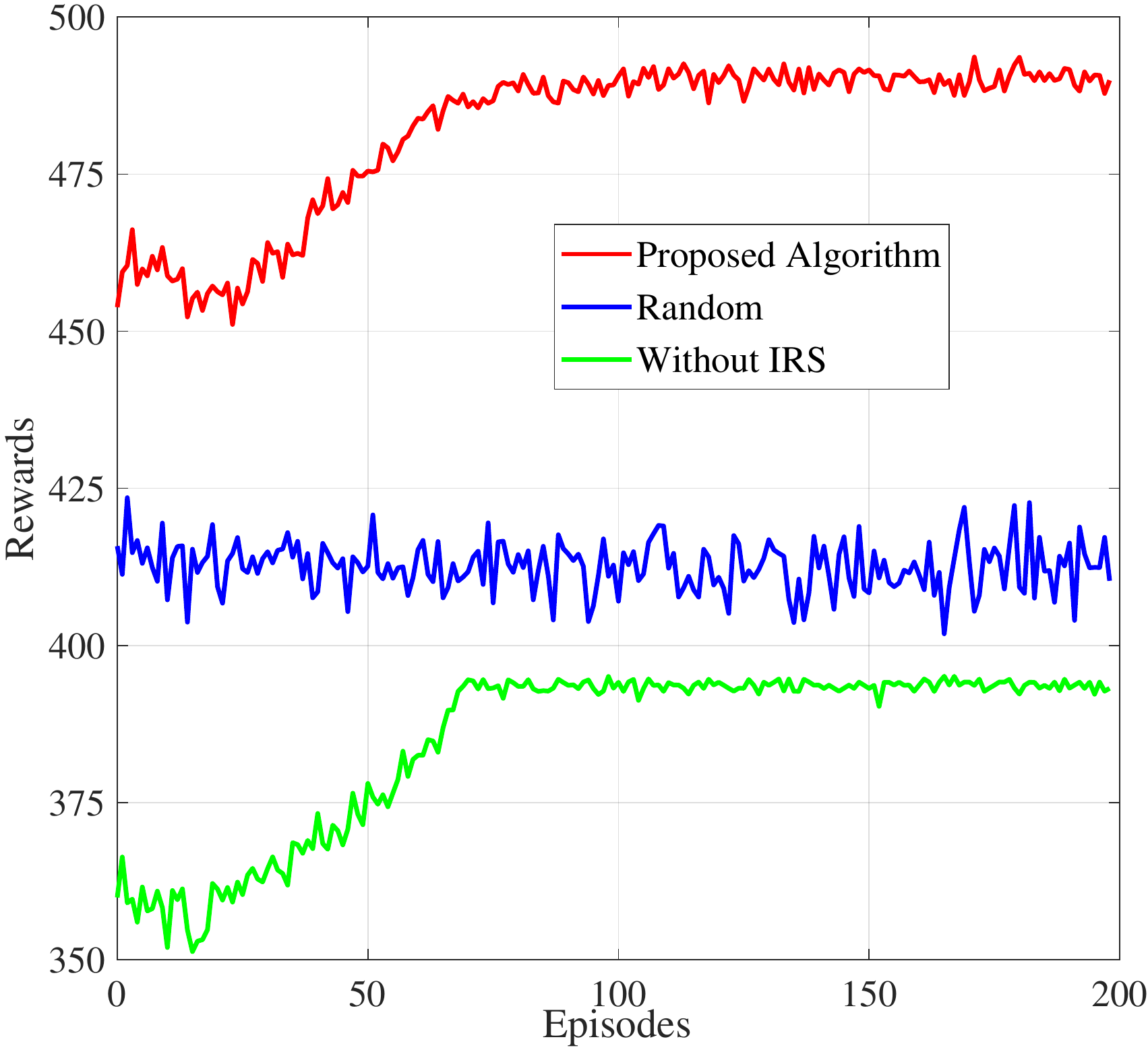}	\label{model}
	\caption{Convergence comparisons of different schemes.}
\end{figure}

\begin{figure*}
\setlength{\abovecaptionskip}{-5pt}
\setlength{\belowcaptionskip}{-10pt}
\centering
\begin{minipage}{5.7cm}
\centering
\includegraphics[width=5.75cm]{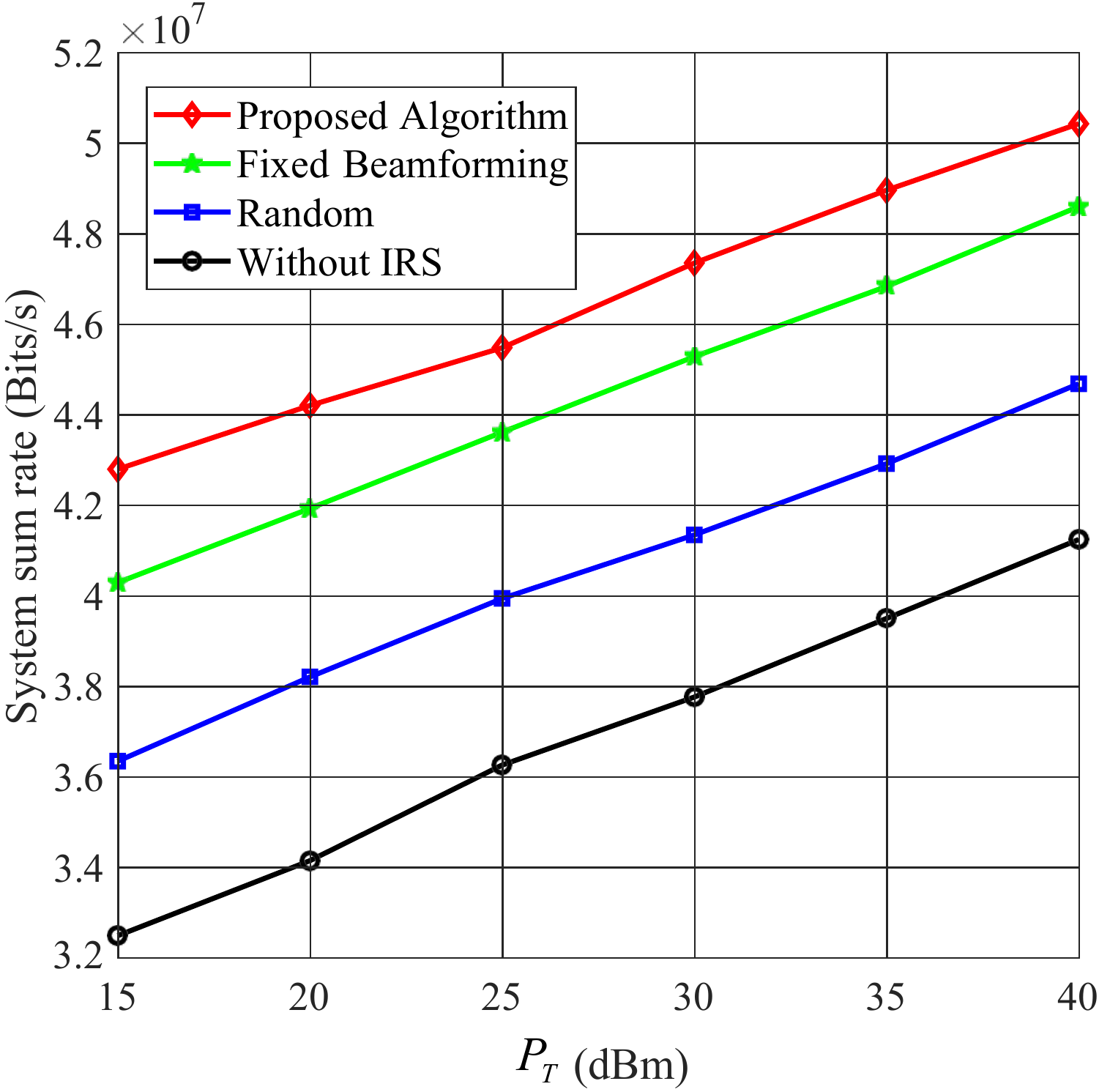}
\caption{System sum rate versus BS transmission power $P_T$ under different schemes.}
\label{figure1}
\end{minipage}
\begin{minipage}{5.7cm}
\centering
\includegraphics[width=5.75cm]{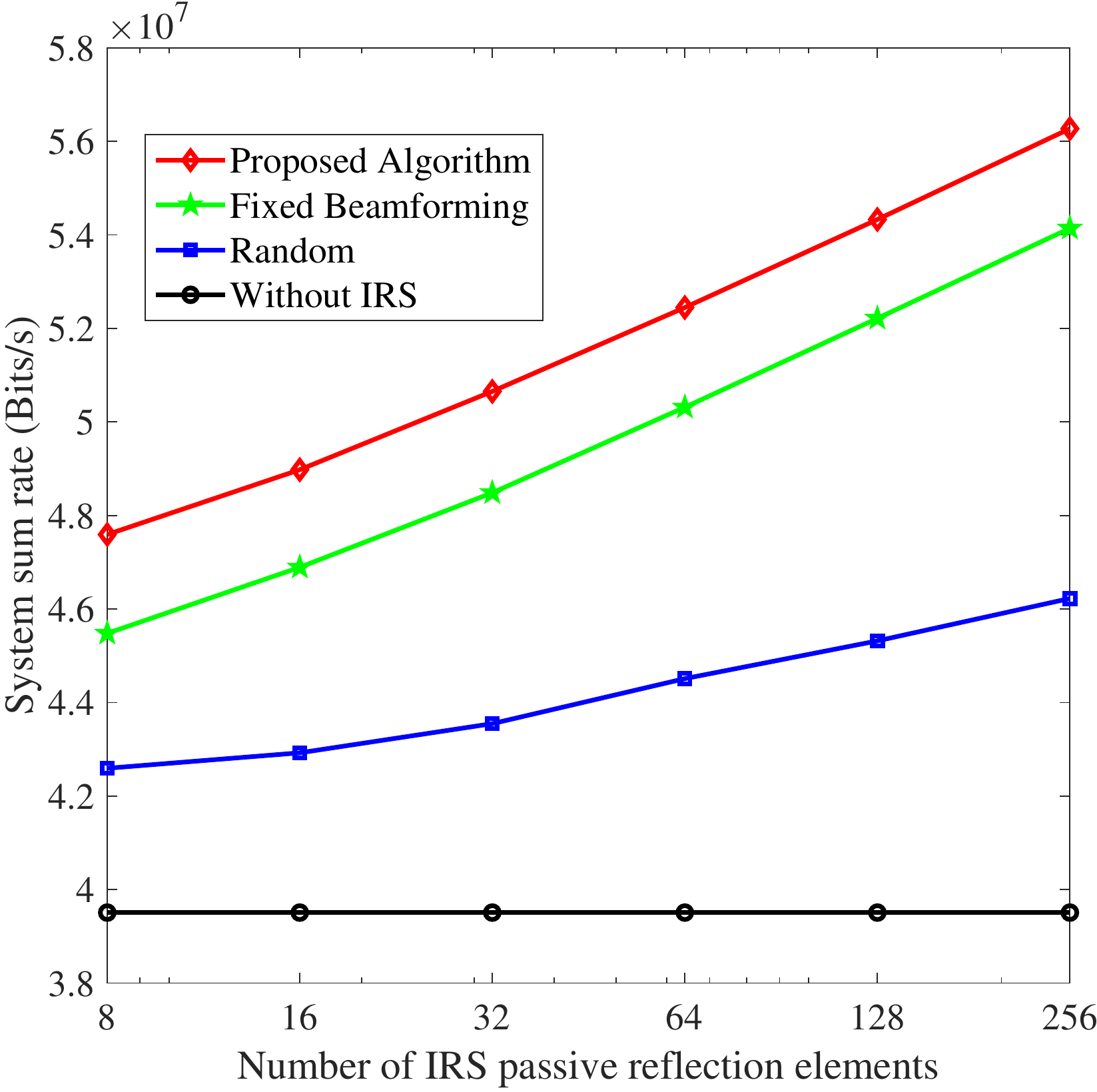}
\caption{System sum rate versus the number of IRS passive reflection elements under different schemes.}
\label{figure2}
\end{minipage}
\begin{minipage}{5.7cm}
\centering
\includegraphics[width=5.75cm]{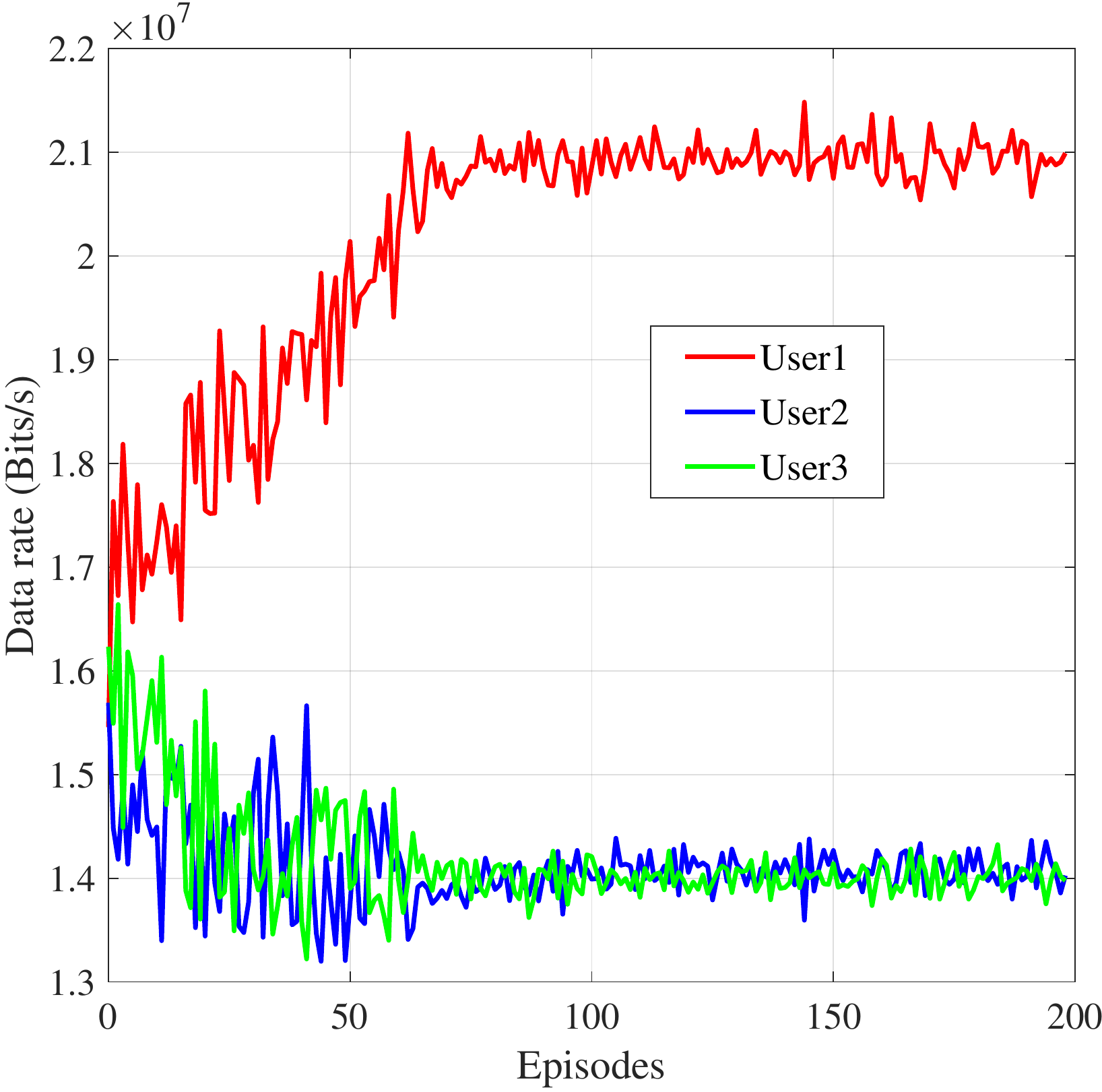}
\caption{Convergence comparison of different users.}
\label{figure3}
\end{minipage}
\end{figure*}

Fig. 4 shows the effection of BS transmission power $P_{T}$ on system sum rate.
For comparison, a fixed beamforming scheme is also introduced.
We set the number of IRS passive reflection elements to 16.
It is observed that the system sum rate increases with the maximum transmission power of the base station.
Our proposed DRL based algorithm has better system sum rate performance than all other three schemes, namely, random selection method, the without IRS method and the fixed beamforming method.

Fig. 5 shows the effection of the number of IRS passive reflection elements on the system sum rate. We set the BS transmission power $P_{T}$ to 35 dBm. 
As shown in Fig. 5 the more number of IRS passive reflection elements the better of system sum rate performance is achieved. 
This is because more reflections units can provide more accurate regulation of the signal phase, and help to achieve higher system sum rate and better communication quality.
It can be seen that when the design complexity of IRS is very high, our proposed algorithm can also obtain better system sum rate than other algorithms.

Fig. 6 shows the convergence performance for different user's data rates. 
It is seen that the data rate of each user tends to converge with the increase of the number of rounds. As the continuous interaction of the environment, the algorithm can learn and adjust the optimization variables to approach the optimal solution.
In practice, the maximum sum data rare is envisioned to be achieved while the users' data rate requirement should be considered. 
It can be seen from Fig. 6 that our proposed algorithm does not just allocate all bandwidth and power to the user with the best channel state for transmission, instead it considers the constraints of the whole network to find the best allocation for each user.

\section{Conclusion}
In this paper, we have investigated the resource allocation problem in IRS-assisted OFDM systems by jointly optimizing the BS beamforming, IRS passive phase shift and channel assignment.
A hybrid MDQN-DDPG framework-based algorithm was proposed to tackle the challenging hybrid discrete and continuous action issue. The system sum rate was maximized while satisfying the minimum transmission rate requirement of user.
Simulation results demonstrated that our proposed algorithm can adjust the action of the agent by observing the immediate reward, and finally make the reward converge to the optimal value. 
The optimal beamforming matrix, IRS phase shift and channel allocation were obtained through the trained agent.

\end{document}